\documentclass[%
reprint,
superscriptaddress,
 amsmath,amssymb,
aps,
prb,
]{revtex4-1}

\usepackage{graphicx}
\usepackage{subcaption} 

\newcommand{\BSS}{Bi$_2$Se$_3$ }
\newcommand{\iton}{$I_{t0}^-$}
\newcommand{\itn}{$I_{t}^-$}
\newcommand{\itop}{$I_{t0}^+$}
\newcommand{\itp}{$I_{t}^+$}

\begin{document}
	
	\title{Experimental detection of surface spin-polarized electron accumulation in topological insulators using scanning tunneling microscopy}
	\author{S. Tyagi}
	\affiliation{Department of Electrical and Computer Engineering, University of Maryland, College Park, MD 20742, USA}
	\author{M. Dreyer}
	\affiliation{Department of Physics, University of Maryland, College Park, MD 20742, USA}
	\author{D. Bowen}
	\author{D.Hinkel}
	\affiliation{Laboratory for Physical Sciences, College Park, MD 20740, USA}
	\author{P.J. Taylor}
	\affiliation{Army Research Laboratory, Adelphi, MD 20783, USA}
	\author{\\ A. L. Friedman}
	\affiliation{Laboratory for Physical Sciences, College Park, MD 20740, USA}
	\author{R.E. Butera}
	\author{ C. Krafft}
	\affiliation{Laboratory for Physical Sciences, College Park, MD 20740, USA}
	\author{I. Mayergoyz}
	\affiliation{Department of Electrical and Computer Engineering, University of Maryland, College Park, MD 20742, USA}
	
	\begin{abstract}
		Spin-momentum locking in the surface mode of topological insulators (TI) leads to the surface accumulation of spin-polarized electrons caused by bias current flows through TI samples. Here, we demonstrate that scanning tunneling microscopy can be used to sense this surface spin-polarized electron accumulation. We present experimental results of this sensing for Sn-doped \BSS samples by employing Fe-coated W tips as well as non-magnetic W tips. We observe a linear increase in the spin-accumulation as a function of bias current through TI samples.  
	\end{abstract}
	\maketitle

\section{Introduction}
	The physical properties of topological insulators (TI) are extensively studied because of their unique features \cite{collTI,ando,qiandzhang7,moore2010birth,qi2010quantum,fu2007topological} as well as their promising applications in the area of spintronics \cite{STT2}. Among these properties is the ninety degree spin-momentum locking \cite{collTI, ando, qiandzhang7, hsieh2009observation} which results in the surface accumulation of spin-polarized electrons when bias currents flow through TI samples. This spin accumulation has been studied theoretically \cite{yazyev2010spin,siu2018spin,peng2016spin,li2016interpreting} as well as using electrical  \cite{STT2,STT1,ando2014electrical,li2014electrical,dankert2015room} and optical experiments \cite{liu2018direct,seifert2018spin}.
	
	 The tunneling behavior of spin-polarized electrons accumulated on TI surfaces \cite{tunn3,tunn1,tunn2,tunn4,tunn5, tunn6, tunn7} is also a subject of ongoing research. For example, tunneling magnetoresistance studies using Ferromagnet-Insulator-TI trilayer devices have been performed to determine the charge-spin conversion efficiencies of TIs \cite{tunn1,tunn3}. In addition, several magnetic memory devices have been proposed based on magnetic tunnel junctions with TIs \cite{tunn3,tunn5, tunn6} which are supposed to have a high tunneling magnetoresistance at room temperature \cite{tunn1}. 
	 
	 Most of these studies were performed using bismuth selenide (Bi$_2$Se$_3$) topological insulators, which has a relatively large bulk energy gap of 0.3 eV as well as an almost ideal single Dirac cone \cite{zhang2009topological}. However, \BSS has high bulk conductivity due to a large amount of Se vacancies \cite{butch2010strong,checkelsky2009quantum}. Various dopings of intrinsic \BSS with elements such as Ca \cite{hor2009p, checkelsky2009quantum, hsieh2009tunable}, Sb \cite{kulbachinskii1999conduction,analytis2010two,tang2019quantum,hong2012ultra}, Sn \cite{dopSn, dopeSn2, ren2012fermi} as well as Se/Te alloying \cite{xiong2012quantum,ren2010large} were used to reduce this bulk conductivity. 
	
	In this work, we present the experimental study of the spin-polarized electron accumulation by using scanning tunneling microscopy (STM).  Previously, STM was used for the measurement of spin-polarized electron tunneling to image magnetization patterns in magnetic samples \cite{fetip1,bode2002magnetization,wiesendanger1994scanning} as well as for the experimental study of spin-Hall effect \cite{xie1,xie3}. The STM technique is extremely local in nature and offers the possibility to correlate spin-polarized electron accumulation with TI surface morphology at the nanoscale.
	 
	 The central idea of our study is to exploit the phenomenon of spin-dependent tunneling to detect surface spin accumulation in TI samples in the presence of bias current flow. We used both Fe-coated W tips as well as non-magnetic W tips. Using Fe-coated W tips, spin accumulation was detected by comparing the tunneling currents measured for opposite directions of bias currents.  In contrast, by using non-magnetic W tips, we detect the spin accumulation in the case of one fixed direction of the bias current. For both these cases, we observe that the spin accumulation is monotonically increased as a linear function of the bias current. 
	
\section{Methods}
	
	We used \BSS samples grown on GaAs substrates using molecular beam epitaxy. These samples were lightly doped with Sn. We used a molecular beam of elemental Sn whose flux was calibrated with a Bayard-Alpert ionization gauge.  The stoichiometry of the used Sn-doped \BSS samples was (Bi$_{0.941}$Sn$_{0.049}$)$_2$Se$_3$. This stoichiometry was determined by calibration of the Sn flux measurements ($\Phi_{\text{Sn}}$) relative to those of pure Bi ($\Phi_{\text{Bi}}$) prior to the growth. The relative Sn concentration was then obtained in the standard manner as $\Phi_{\text{Sn}}/ (\Phi_{\text{Bi}} + \Phi_{\text{Sn}})$.  Evidence suggests that Sn-doping can suppress the conductivity of the sample bulk and enhance in this way the portion of the bias current carried by the topologically protected conducting surface mode \cite{tyagi2020study, dopSn,dopeSn2,ren2012fermi}. Since \BSS has a relatively large band gap of 0.3 eV and a single almost ideal Dirac cone, Sn-doped \BSS samples are promising platforms to observe spin-polarized electron accumulation due to TI surface states. 
	
	The W tips used in our STM experiments were prepared through electrochemical etching. The Fe-coated W tips were prepared by electron-beam deposition of approximately 20 nm of iron on clean, etched W tips. For this Fe film thickness, we expect the magnetization to be predominantly tangential to the tip surface in the tip apex area.

\begin{figure}[h!]
	\includegraphics[width=0.9\columnwidth]{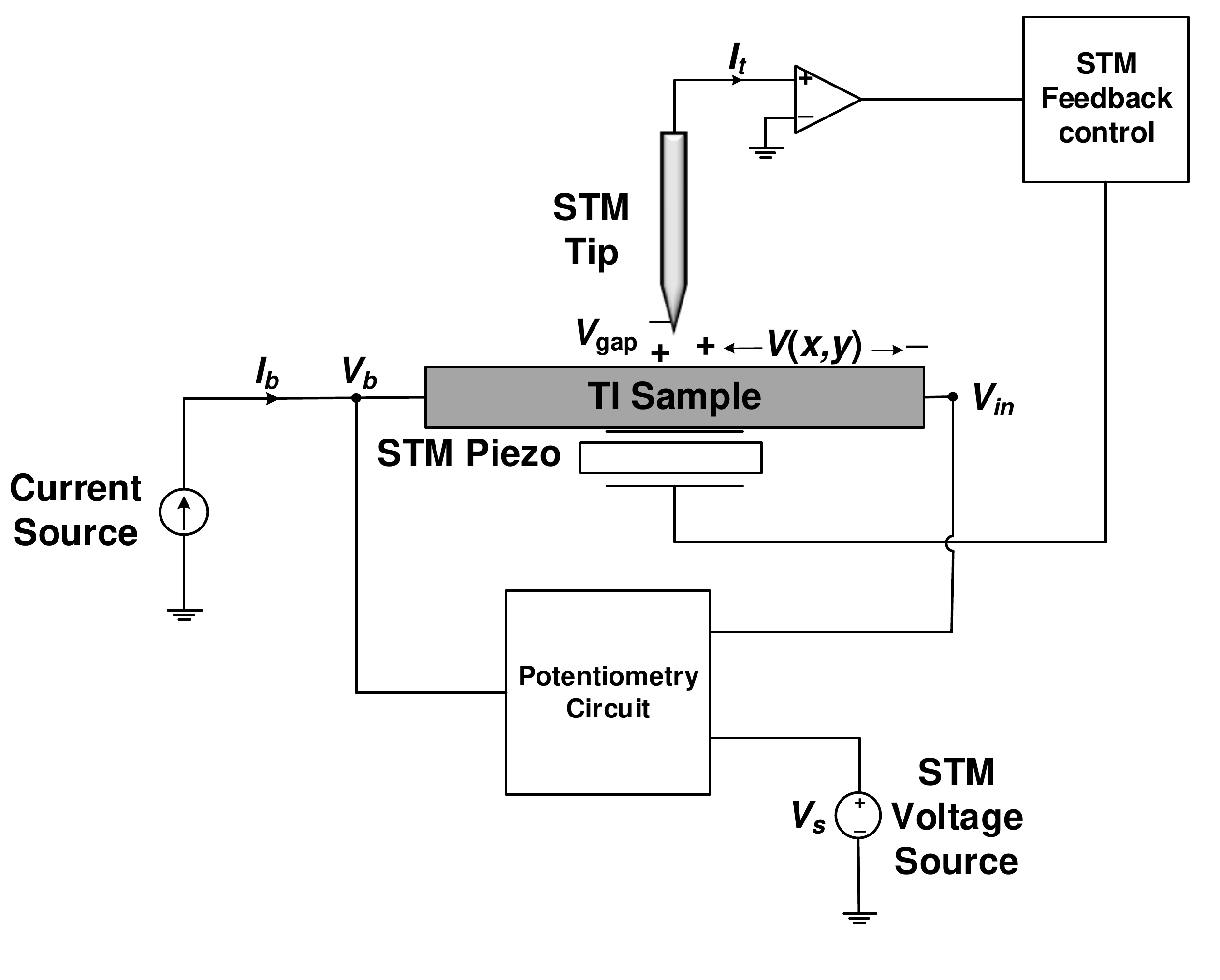}
	\caption{Schematic of STM set up. A current source provides the desired bias current through the TI sample, while the potentiometry circuit ensures that $V_{gap}=V_s$ for varying bias currents.} 
	\label{Pot}
\end{figure}

	The STM experiments were performed at room temperature and in ultra-high vacuum (UHV) by using a two chamber Omicron UHV-STM system. A schematic of the experimental set-up is shown in Figure \ref{Pot}. The current source shown in this figure was used to provide the desired bias current through the sample, while the voltage source was used to achieve the desired gap voltage between the STM tip and the sample. The gap voltage was maintained to be constant in the presence of bias currents by using a previously developed potentiometry technique \cite{xie1}. This technique compensates for the voltage drop $V(x,y)$ across the sample caused by bias current flow. 

\section{Results and Analysis}
\begin{figure}[h!]
	\includegraphics[width=0.9\columnwidth]{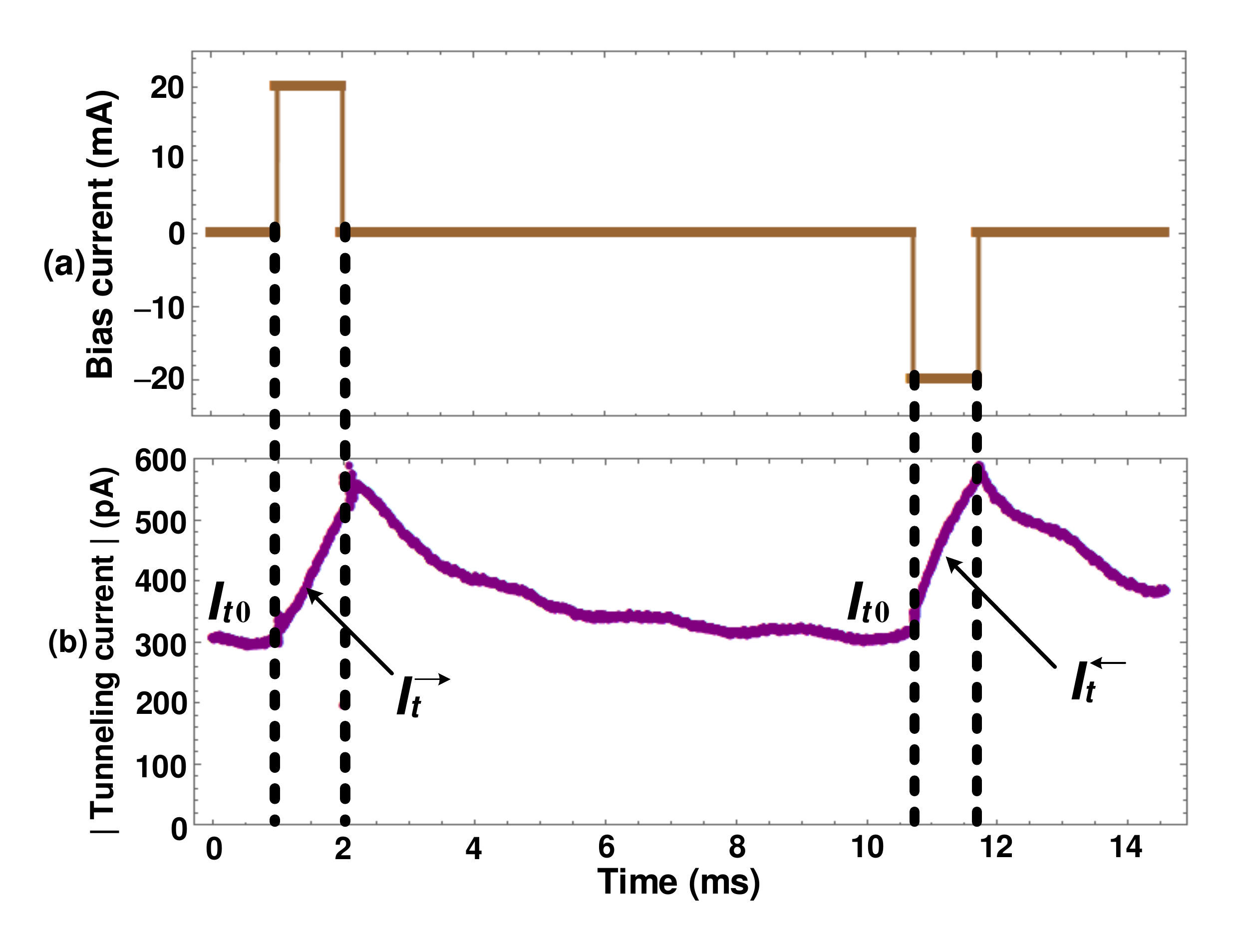}
	\caption{Graphical illustration of experiments performed to measure difference in tunneling for opposite directions of spin accumulations on the surface of Sn-doped \BSS by using Fe-coated W tips. (a) Bias-currents of opposite directions pulsed through the sample for the same gap voltage, (b) the tunneling current magnitude measured as functions of time}  
	\label{FeExpt}
\end{figure}

Figure \ref{FeExpt} shows the procedure used in our experiments for opposite bias current directions. We started with zero bias current through the sample, a chosen value of negative gap voltage $V_{gap}^-$ and a set-point tunneling current of magnitude $I_{t0}$. For a negative gap voltage, electrons tunnel from the sample to the tip. Then the STM feedback was turned off, and after a short delay (a few milliseconds) a bias current was pulsed through the sample. The chosen tunneling voltage was maintained in the presence of the bias current pulse by the potentiometry circuit (Figure 1). The pulsed bias current resulted in a monotonic increase in the magnitude of tunneling current, $I_t^\leftarrow$. After the end of the bias current pulse, the tunneling current gradually reduced back to its initial value $I_{t0}$.  Subsequently, for the same polarity of the tunneling voltage, a bias current of the same magnitude and width but the opposite direction was pulsed through the sample, and the corresponding tunneling current magnitude $I_t^{\rightarrow}$ was measured. The two tunneling current magnitudes $I_t^{\leftarrow}$ and $I_t^{\rightarrow}$ were then compared as functions of time.

\begin{figure}[h!]
	\includegraphics[width=0.9\columnwidth]{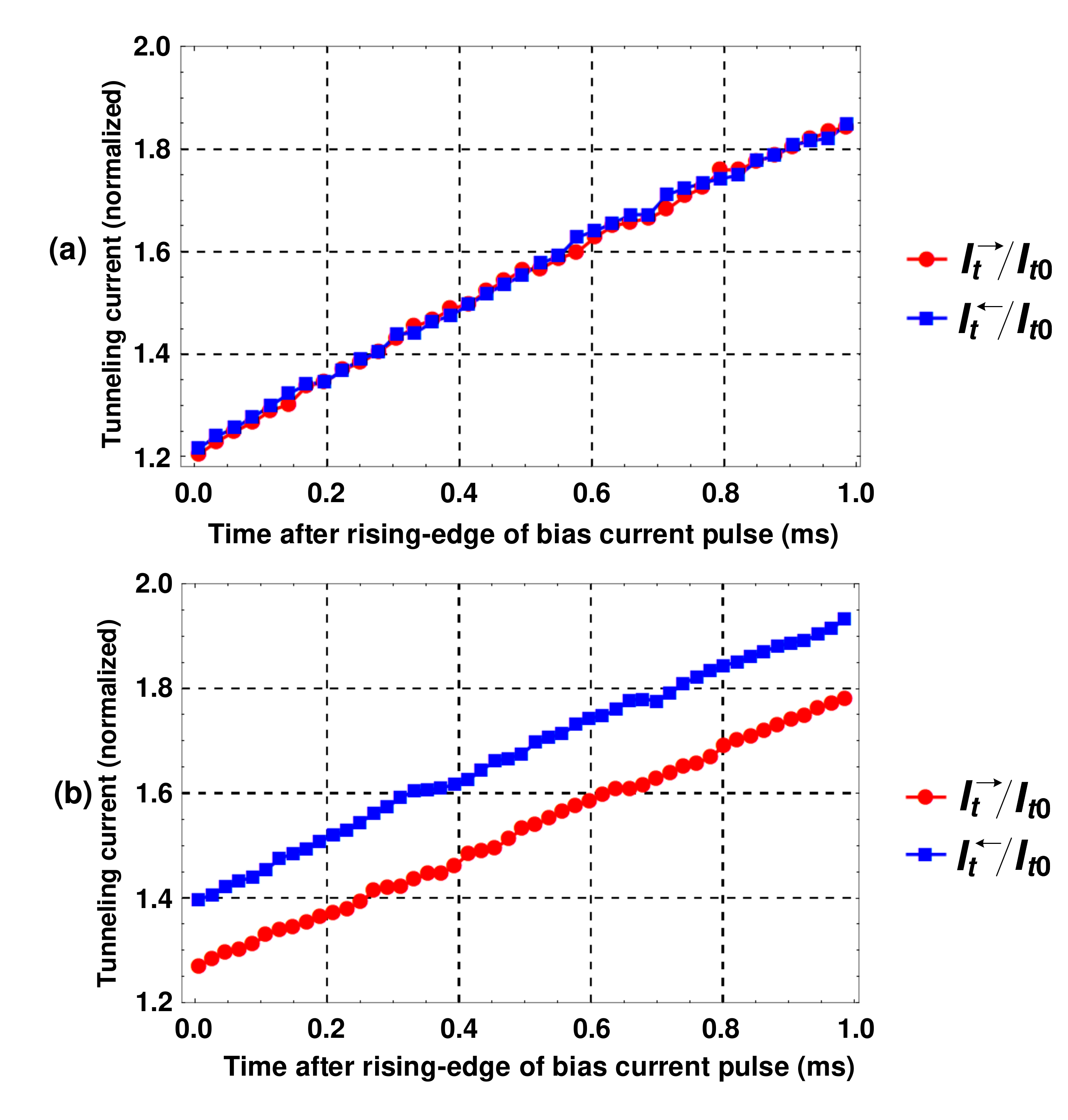}
	\caption{Normalized tunneling currents $I_t^{\leftarrow}/I_{t0}$ and $I_t^{\rightarrow}/I_{t0}$ obtained from the measurements performed with (a) W tips (b) Fe-coated tips.}  
	\label{f5}	
\end{figure}

It is evident from Figure \ref{FeExpt} that the pulse bias currents lead to a monotonic increase in tunneling current. This is due Joule heating of the sample, which leads to its thermal expansion and, consequently to a reduction in the tunneling gap. This thermal contribution obscures the tunneling current detection of spin-polarized electrons on TI sample surface. Therefore, it is imperative to maintain identical thermal conditions at the beginning of each pulse. This was done by sufficiently separating the bias current pulses in time to achieve the same initial tunneling current $I_{t0}$ (and same tunneling gap) before bias currents were pulsed through the sample. This ensures that thermal expansions caused by bias-currents were identical for all current pulses of the same magnitude. Indeed, as demonstrated in Figure ~\ref{f5}(a), when the measurements in the above experiment were performed using non -magnetic clean W tips, it was observed that the tunneling currents $I_t^{\leftarrow}$ and $I_t^{\rightarrow}$ were practically identical. This implies the identical thermal expansions for bias currents of opposite directions. This also implies the identical tunneling voltages for opposite directions of bias currents achieved due to the potentiometry technique. 

Figure ~\ref{f5}(b) shows an example of the tunneling currents measured when the above experiment was performed with an Fe-coated (magnetic) W tip. It is evident from this figure that these normalized tunneling currents are not identical, which can be understood as a result of spin-dependent tunneling of spin-polarized electrons from the TI sample to the Fe-coated W tip. Indeed, as discussed in the previous paragraph, for pulses of bias currents of opposite directions, the tunneling takes place under identical conditions. The only difference is the opposite electron spin directions on the sample surface for opposite bias-current directions as a result of spin-momentum locking. These opposite electron spin directions may affect the tunneling currents only due to spin-dependent density of states in Fe-coated W tips \cite{wiesendanger1994scanning}. There is no such difference in these tunneling currents when measured by using non-magnetic W tips. 

\begin{figure}[h!]
	\begin{subfigure}{.5\textwidth}
		\includegraphics[width=0.9\columnwidth]{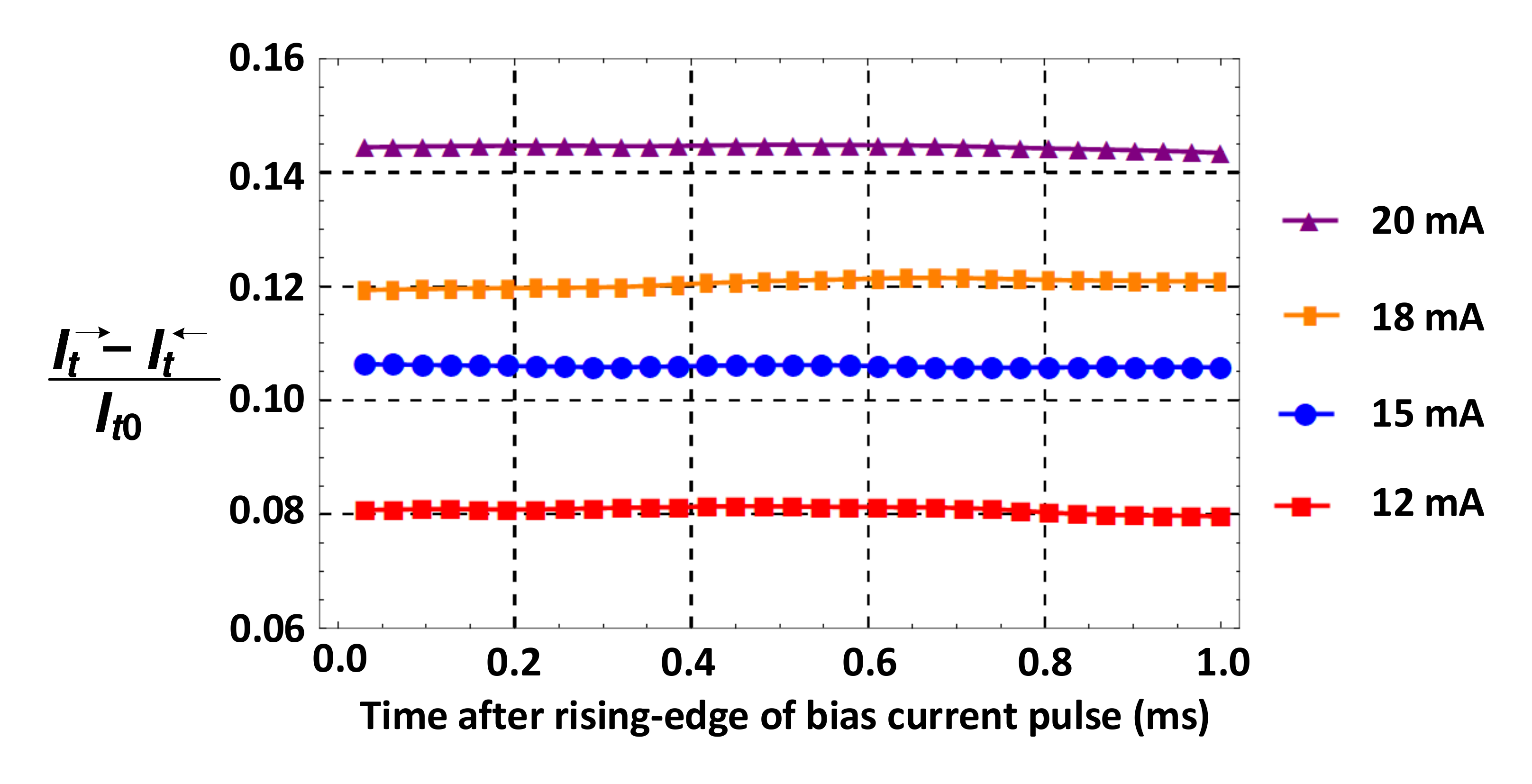} 
		\subcaption{}
	\end{subfigure}
\begin{subfigure}{.5\textwidth}
	\includegraphics[width=0.9\columnwidth]{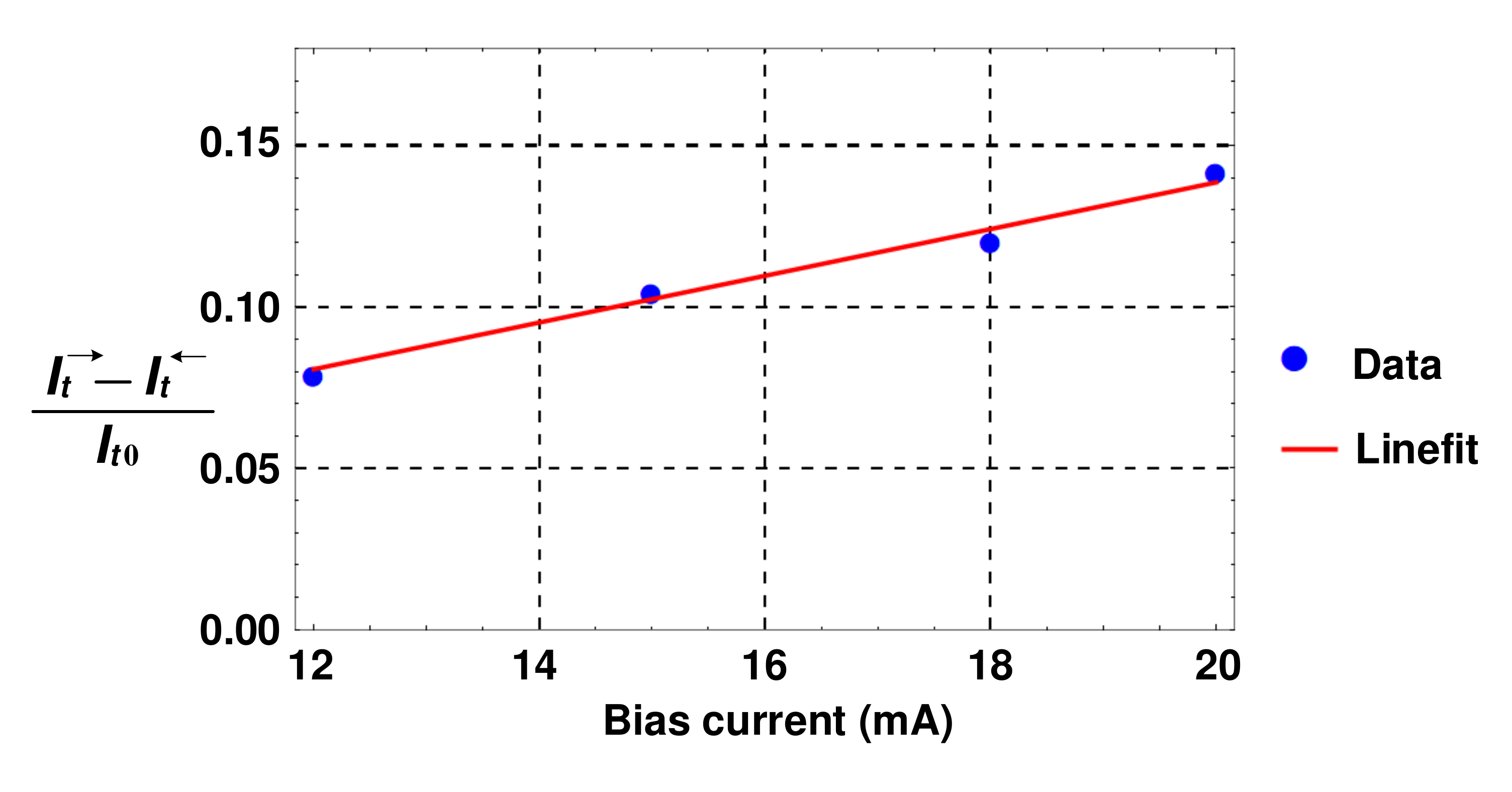} 
	\subcaption{}
\end{subfigure}
	\caption{(a) Normalized difference in tunneling currents for opposite directions of bias-currents in the case of Fe-coated W-tips (b) The linear increase of normalized difference in tunneling currents for the monotonic increase of pulsed bias currents.} 
	 \label{f6}
\end{figure}

Figure \ref{f6}(a) presents the tunneling current differences for various magnitudes of pulsed bias currents. The differences in tunneling currents are stabilized in time, and do not have any thermal contributions. Consequently, the observed (normalized) difference $\left(I_t^{\leftarrow} - I_t^{\rightarrow}\right)/I_{t0}$ in the tunneling currents for opposite directions of pulsed bias currents can be viewed as a measure of surface spin-polarized electron accumulation. Additionally,  the above differences are increased monotonically as functions of the bias current magnitude. As illustrated by Figure \ref{f6}(b), this monotonic increase is linear in nature, which may be attributed to the greater accumulation of spin-polarized electrons on the sample surface in the case of larger bias currents. The above observation is consistent with theoretical calculations of the charge-spin conversion efficiency \cite{zhang2016conversion} for TIs, which suggest the spin-accumulation linearly depends on the magnitude of the bias current.  

\begin{figure}[h!]
		\includegraphics[width=0.8\columnwidth]{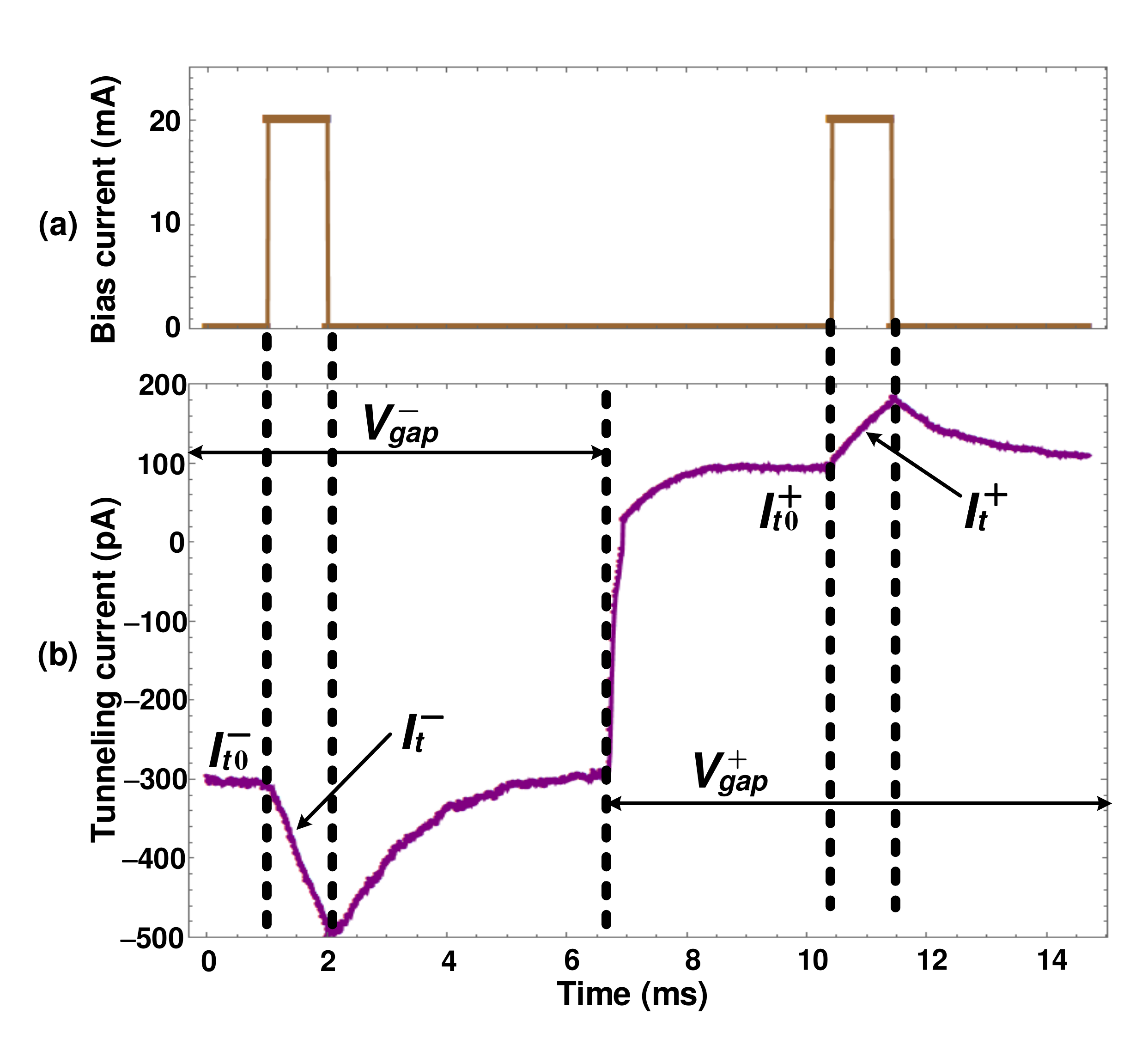}
		\caption{Graphical illustration of the experiment to detect surface spin accumulation for the same direction of bias currents with clean W-tips, for $V_{gap}^+ = 0.1$ V and $V_{gap}^- = -0.1$ V,\iton = $-300$ pA and \itop = $112$ pA. (a) Same magnitude and polarity bias current pulses applied, (b) Tunneling currents measured for opposite values of gap voltages.}  
		\label{WtipExp}
	\end{figure}

Figure \ref{WtipExp} illustrates the measurement of spin-polarized electron accumulation for one fixed direction of bias current using clean W tips. The measurements began with zero bias current and a chosen value of negative tunneling voltage $V_{gap}^-$ and the fixed current $I_{t0}^-$ of electrons tunneling from the sample to the tip. Then, the STM feedback was turned off and after a brief delay ($\sim$5ms) some bias current was pulsed through the sample, and the corresponding tunneling current, $I_t^-$ of spin-polarized electrons was measured as a function of time. After the end of the bias current pulse, the tunneling current gradually reduced back to its initial value \iton. Subsequently, the same procedure was repeated for the tunneling voltage $V_{gap}^+$ of positive polarity but the same magnitude as $V_{gap}^-$. The corresponding tunneling currents $I_{t0}^+$ and $I_{t}^+$ of non-spin polarized electrons from the tip to the sampled were measured. 

\begin{figure}[h!]
	\includegraphics[width=\columnwidth]{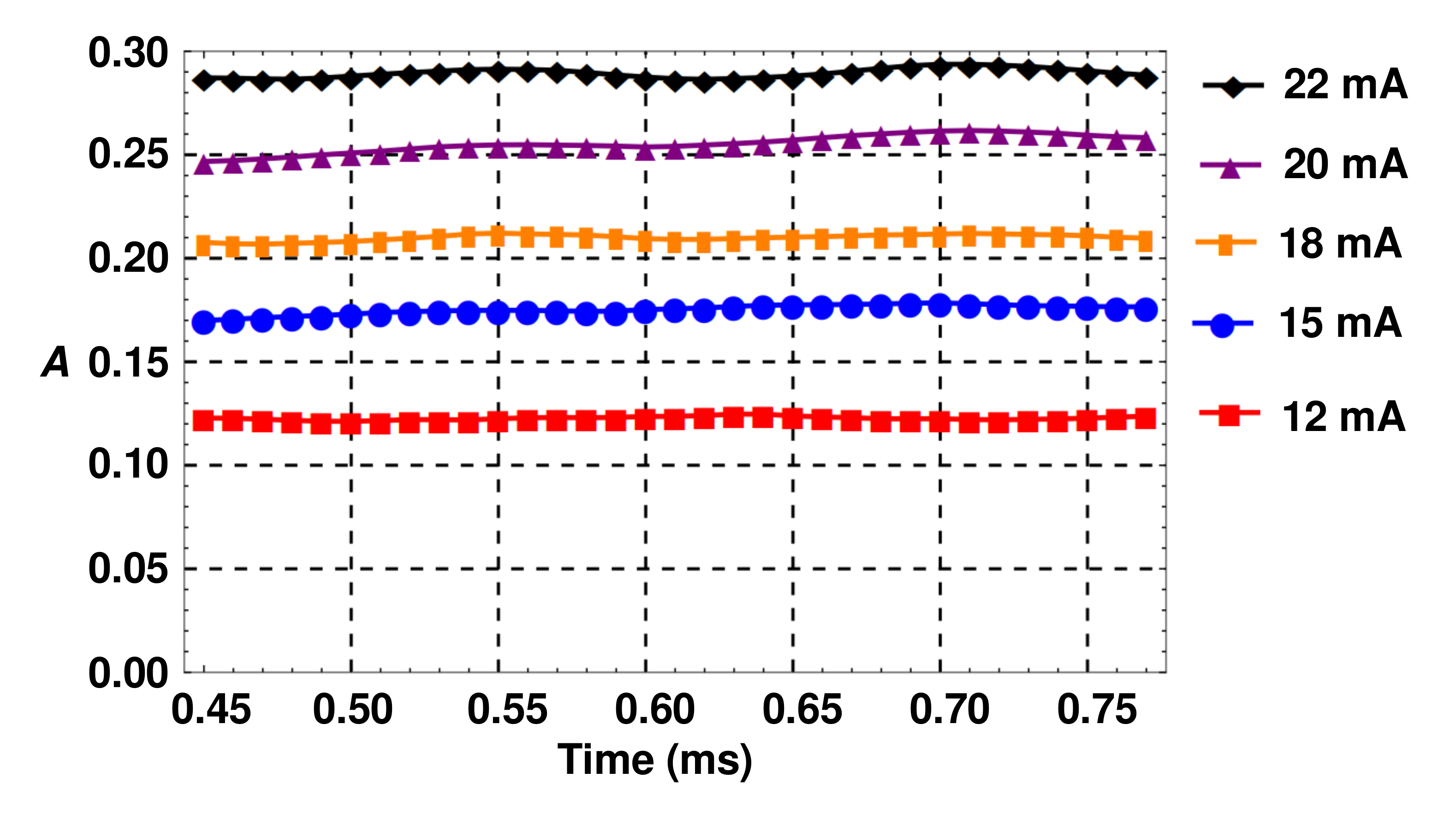}
	\caption{The values of $A$ as functions of time measured with clean W tips for different values of bias-currents.} 
	\label{WtipRes} 
\end{figure}

The results of the performed experiments are presented in Figure \ref{WtipRes} for various magnitudes of rectangular bias current pulses in terms of the following quantity $A$:
	\begin{equation}
		A = \frac{\frac{I_{t}^-}{I_{t0}^-} - \frac{I_{t}^+}{I_{t0}^+}}{\frac{I_{t}^-}{I_{t0}^-} + \frac{I_{t}^+}{I_{t0}^+}}.
		\label{eqA}
	\end{equation}

The choice of $A$ was made to eliminate the thermal effects on the tunneling currents caused by the pulses of bias currents. At constant tunneling voltages, the change in tunneling currents during bias current pulses can be affected by two factors: 1) continuous reductions in tunneling-gaps due to sample thermal expansion, and 2) surface accumulation of spin-polarized electrons.  Figure~\ref{WtipRes} reveals that the quantity $A$ becomes constant in time for rectangular bias current pulses of fixed magnitudes. The constant in time values of $A$ for fixed bias currents suggest that the thermal effects do not affect $A$. It can be argued that this occurs because exponential increases of tunneling currents \itn and \itp as a function of time due to thermal expansion are canceled in formula (1). Thus, the above quantity $A$ provides a measure of the accumulation of spin-polarized electrons on the surface of \BSS samples in the presence of bias currents.

\begin{figure}
	\includegraphics[width=0.8\columnwidth]{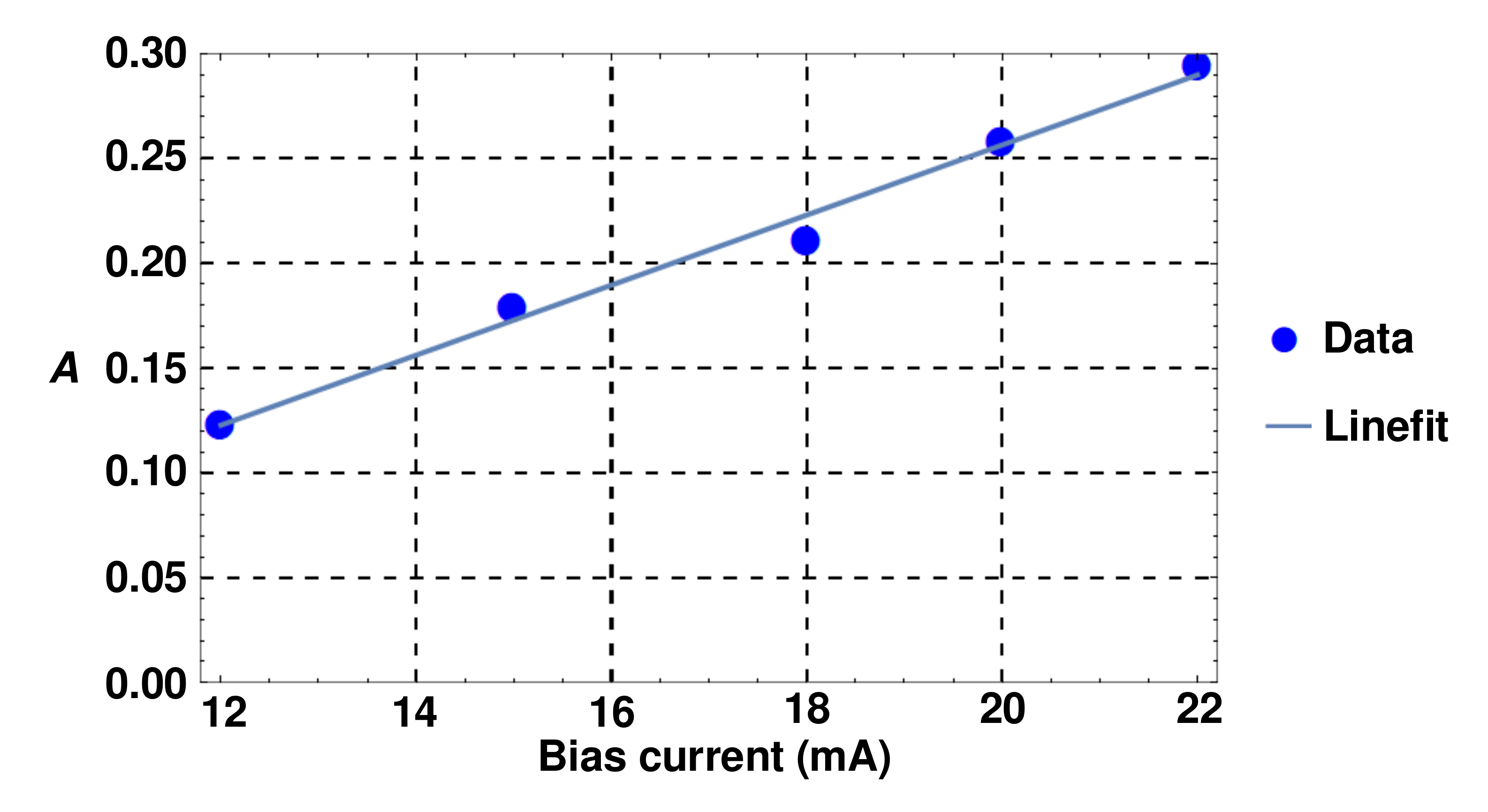}
	\caption{The linear increase of $A$  as a function of bias current.} \label{AvsIbW}
\end{figure}

Furthermore,  the measured $A$ increases monotonically as a function of bias current because of the enhancement in surface spin-polarized electron accumulations due to spin-momentum locking. This interpretation is supported by the fact that the tunneling of spin-polarized electrons is physically different from the tunneling of non spin-polarized electrons. Indeed, the tunneling of spin-polarized electrons must be described by the Schrödinger equation with the spin-orbit interaction (Dresselhaus \cite{dresselhaus1955spin}) terms in the Hamiltonians, while no such terms are used in the study of non spin-polarized electrons tunneling. The theoretical study of spin-polarized electrons by using Hamiltonians with spin-orbit terms was performed previously by V.I. Perel and his coauthors \cite{perel2003spin,tarasenko2004plane}. It has also been demonstrated using STM that W-tips can detect differences between in-plane and out-of-plane magnetization in magnetic films \cite{bode2002magnetization}. This suggests the spin-sensitivity of nonmagnetic W-tips. Hence,  different values of $A$ for different bias currents reflect different levels of surface spin-polarized electron accumulations caused by these bias currents. This conclusion is consistent with the linear increase in $A$ as a function of bias-current through the TI sample, as shown in Fig.~\ref{AvsIbW}. 

\section{Conclusion}
The surface spin-polarized electron accumulation in topological insulators caused by a bias current flow is studied by using STM experiments. The results of this study are presented for Sn-doped \BSS samples by employing Fe-coated W tips as well as W tips. The current flow through samples results in their heating and thermal expansion, and these effects obscure the influence of surface spin-polarized electron accumulations on tunneling. It is demonstrated that these thermal effects can be properly filtered out for both W and Fe-coated W tips. It is also observed that the increase in bias-current magnitude leads to monotonic (linear) increase of spin polarized electrons accumulation on the surface of TI samples. The presented technique can be extended to study the correlations of tunneling behavior of these Dirac electrons with surface morphology of topological insulators at the nanoscale level. Furthermore, it is apparent that TIs may serve as unique testbeds for the characterization of iron-coated tips as well as for the study of spin-polarized tunneling.

\bibliography{References_check}
\end{document}